\documentclass[10pt,twocolumn,superscriptaddress,preprintnumbers,showpacs]{revtex4}
\usepackage{epsf} 
\usepackage{amsmath}
\usepackage{amssymb}
\usepackage{epsfig}
\usepackage{latexsym}
\usepackage{amsfonts}
\usepackage{graphicx}%
\usepackage{varioref}
\usepackage{ifthen}
\begin{document}
\parindent 0mm 
\parindent 0mm 
\setlength{\parskip}{\baselineskip}
\thispagestyle{empty}
\pagenumbering{arabic} 
\setcounter{page}{1}
\mbox{ }
\preprint{UCT-TP-276/09}
\title	{QED vacuum fluctuations and induced electric dipole moment of the neutron}
\vspace{.1cm}
\author{C. A. Dominguez} 
\affiliation{Centre for Theoretical Physics and Astrophysics, University of
Cape Town, Rondebosch 7700}
\affiliation{Department of Physics, Stellenbosch University, Stellenbosch 7600, South Africa}

\author {H. Falomir} 
\affiliation{IFLP - CONICET, Universidad Nacional de La Plata, Argentina}

\author{M. Ipinza} 
\affiliation{Facultad de F\'{\i}sica, Pontificia Universidad Cat\'{o}lica de Chile, Casilla 306, Santiago 22, Chile}

\author{S. Kohler}
\affiliation{Department of Physics, Stellenbosch University, Stellenbosch 7600, South Africa}

\author{M. Loewe} 
\affiliation{Facultad de F\'{\i}sica, Pontificia Universidad Cat\'{o}lica de Chile, Casilla 306, Santiago 22, Chile}

\author{J. C. Rojas} 
\affiliation{Departamento de F\'{i}sica,  Universidad Cat\'{o}lica del Norte, Casilla 1280, Antofagasta, Chile}

\date{\today}

\begin{abstract}
\noindent
Quantum fluctuations in the QED vacuum generate non-linear effects, such as peculiar induced electromagnetic fields. In particular, we show here that an electrically neutral particle, possessing a magnetic dipole moment, develops an induced electric dipole-type moment with unusual angular dependence, when immersed in a quasistatic, constant external electric field.
The calculation of this effect is done in the framework of the Euler-Heisenberg effective QED Lagrangian, corresponding to the weak field asymptotic expansion   of the effective action to one-loop order. It is argued that the neutron might be a good candidate to probe this signal of non-linearity in QED. 
\end{abstract}
\maketitle
\noindent
It has been known since the pioneering work of Euler and Heisenberg \cite{H-E}-\cite{REVS} that quantum fluctuations in the QED vacuum induce non-linearity.  This happens in spite of QED being a linear theory at the Lagrangian level. Given the extreme smallness of these induced non-linear effects, this subject has remained in relative obscurity for many years. Recent renewed interest in this field has been motivated in part by  
the steady increase in the intensity of modern lasers \cite{LASERS} with associated peak electric fields reaching $10^{14} $ V/m,  and envisaged fields of $10^{15} - 10^{16}$ V/m in the near future. Such intense fields are approaching the critical value $E_c \simeq 10^{18}$ V/m beyond which unusual properties of the QED vacuum are expected to manifest themselves. Among these properties one has vacuum birefringence, and non-linear Compton scattering \cite{Tom}.
In addition to such time-dependent non-linear phenomena, classical charge/current sources immersed in external quasistatic fields also generate unexpected electromagnetic field configurations induced by QED vacuum fluctuations. In fact, we have  shown recently \cite{NLQED1} (see also \cite{CAN}) that an electrically charged (non-magnetic) particle immersed in a constant external magnetic field generates several electric and magnetic multipoles with unusual angular dependences. Among these there is an electric multipole term, independent of the external magnetic field, which corresponds to the correction to the Coulomb field due to quantum fluctuations in the QED vacuum. More importantly, though, the leading term of the induced magnetic field is due to an induced magnetic dipole moment. For accessible values of the relevant parameters, this induced effect is comparable to the magnetic dipole moment of the nucleon.

Given the intrinsic relationship between electric and magnetic phenomena in Electrodynamics, the reciprocal effect  is to be expected, i.e. that a magnetic dipole immersed in a constant electric field might generate an electric dipole moment.
In this paper we show that this is in fact the case, and argue that the neutron might serve as a probe of this induced non-linear effect in QED. 
While, in principle, these non-linear effects could be studied within the full QED theory, their complexity calls for a simplifying approach such as  that  of a corresponding  effective theory. For instance, looking at photon-photon scattering mediated by fermionic loops, and in the kinematical regime where the photon momenta are much smaller than the fermion rest masses, there follows an effective QED action. In the case of quasiatatic electromagnetic fields, the first non-linear term in the weak field expansion is the Euler-Heisenberg Lagrangian \cite{H-E}-\cite{REVS}
%Eq.1
\begin{equation}
        \mathcal{L}^{(1)} =  \zeta \left(
      4 \mathcal{F}^2 + 7 \mathcal{G}^2 \right) +... \;
\end{equation}
where the omitted terms are of higher order in the expansion parameter $\zeta$, which in SI units is given by
%Eq.2
\begin{equation}
\zeta = \frac{ 2 \alpha^2 \varepsilon_0^2 \hbar^3}{45 m_e^4 c^5} \simeq \;\, 1.3 \times 10 ^{- 52}\; \frac{\mbox{J\, m}}{\mbox{V}^4} \;,
\end{equation}
with $\alpha = e^2/(4 \pi \varepsilon_0 \hbar c)$ the electromagnetic fine structure constant, and $c$ the speed of light. 
The invariants $\mathcal{F}$ and $\mathcal{G}$ are defined as
%Eq.3
\begin{equation}
    \mathcal{F}= \frac{1}{2} \left( \mathbf{E}^2 - c^2 \;\mathbf{B}^2 \right)
    = - \frac{1}{4} F_{\mu \nu} F^{\mu \nu}\,,
\end{equation}
%Eq.4
\begin{equation}    
    \quad \mathcal{G}= c \;\mathbf{E} \cdot \mathbf{B}
    = - \frac{1}{4} F_{\mu \nu} \widetilde{F}^{\mu \nu} \,,
\end{equation}
with $F_{\mu \nu} = \partial_\mu A_\nu - \partial_\nu A_\mu$ and $\widetilde{F}^{\mu \nu} = \frac{1}{2} \epsilon^{\mu\nu\rho\sigma} F_{\rho\sigma}$. 
Clearly, it is possible to use a different, dimensionless expansion parameter of the QED effective action in terms of e.g. a critical electric or magnetic field for the onset of non-linearity. This critical field  is roughly given by that needed to produce an electron-positron pair from the vacuum in a length scale of a Compton wavelength, i.e.
$E_c= m_e^2 c^3/\hbar e \simeq 1.3 \times 10^{18} \; \mbox{Volt/m}$,
where $m_e$ and $e$ are the mass and charge of the electron, respectively. In this case, 
$\zeta = \kappa (m_e/E_c)^4$, where $\kappa \equiv 2 \alpha^2 \epsilon_0^2 \, c^7/(45 e^4 \hbar)$, or using natural units ($\hbar = c = 1$) one has $\kappa = 1/(360 \pi^2)$. However, in the sequel we use $\zeta$ as the expansion parameter.
The constitutive equations which follow in the non-linear effective theory are very different from the standard ones: $\mathbf{D} = \varepsilon_0 \mathbf{E} + \mathbf{P}$, and $\mathbf{H} = \frac{\mathbf{B}}{\mu_0} - \mathbf{M}$, where $\mathbf{P} = \mathbf{M} \equiv 0$, if quantum vacuum fluctuations are ignored. In fact, from the full Lagrangian, Eq.(1),
it follows that
%Eq.5
\begin{equation}
      \mathbf{D} = \frac{\partial \mathcal{L}}{\partial \mathbf{E}}= \left( \frac{\partial \mathcal{L}}{\partial \mathcal{F}} \right) \mathbf{E}
      + \left( \frac{\partial \mathcal{L}}{\partial \mathcal{G}} \right) c \mathbf{B}
      \,,
\end{equation}
%Eq.6
\begin{equation} 
      \mathbf{H}= - \frac{\partial \mathcal{L}}{\partial \mathbf{B}} =\left( \frac{\partial \mathcal{L}}{\partial \mathcal{F}} \right)  c^2\, \mathbf{B}
      - \left( \frac{\partial \mathcal{L}}{\partial \mathcal{G}} \right) c \,\mathbf{E}
     \,.
\end{equation}
These equations imply the well known non-linear relations
$\mathbf{P} = 2 \zeta (4\, \mathcal{F}\, \mathbf{E} + 7 \,c \,\mathcal{G} \,\mathbf{B})\;,$ and
$\mathbf{M} = 2\, \zeta (- 4 \,c^2 \,\mathcal{F}\, \mathbf{B} + 7\, c \,\mathcal{G}\, \mathbf{E})\;.$
In terms of the fields {\bf D} and {\bf H} the Euler-Lagrange equations of motion  reduce to
the linear Maxwell equations with all the effects of non-linearity  contained in the constitutive equations, Eqs.(5) and (6).
Therefore, the equations for $\mathbf{D}$ and $\mathbf{H}$ can be solved as in the usual linear theory, i.e. 
$\nabla \cdot \mathbf{D}=j_0$,
$- \partial \mathbf{D}/\partial t+\nabla\times \mathbf{H} = \mathbf{j}$,
$\nabla \cdot \mathbf{B}=0$,
$\nabla\times \mathbf{E} +
      \partial \mathbf{B}/\partial t=0$.
The relations between the fields  $\mathbf{D}$ and  $\mathbf{H}$ and the electromagnetic intensities $\mathbf{E}$ and $\mathbf{B}$, Eqs.(5) and (6), can be easily inverted to leading order in $\zeta$ (see \cite{NLQED1}).
As shown in \cite{NLQED1} it is possible to obtain a general analytical solution to the inhomogeneous  equations for $\mathbf{D}(\mathbf{x})$ and $\mathbf{H}(\mathbf{x})$  in terms of the gradient of a scalar function $\phi(\mathbf{x})$, and the curl of a vector function $\mathbf{K}(\mathbf{x})$, both of order $\cal{O}$$(\zeta)$, and calculable in terms of the external electric/magnetic sources. One first writes \cite{extra1}
%Eq.7
\begin{equation}
      \mathbf{D(\mathbf{x})}=\mathbf{D}(\mathbf{x})_M + \nabla \times \mathbf{K}(\mathbf{x})\,,
\end{equation}
%Eq.8
\begin{equation}
        \mathbf{H}(\mathbf{x})= \mathbf{H}(\mathbf{x})_M + \nabla \phi(\mathbf{x})\,,
\end{equation}
where $\mathbf{D}(\mathbf{x})_M$ and $\mathbf{H}(\mathbf{x})_M$ are the solutions in the linear theory which satisfy $\nabla \times \mathbf{D}_M =0$, and $\nabla \cdot \mathbf{H}_M =0$. 
The vectors $\nabla \times \mathbf{K}(\mathbf{x})$, and  $\nabla \phi(\mathbf{x})$ can be obtained using the remaining homogeneous equations with the result
%Eq.9
\begin{equation}
\nabla \phi (\mathbf{x}) = \frac{\zeta \, c}{2\pi \varepsilon_0}  \nabla_{\mathbf{x}} \int
\frac{d^3y}{\left|\mathbf{x} - \mathbf{y} \right|}  \nabla_{\mathbf{y}} \cdot [ 7 \mathcal{G} \mathbf{D} 
-  \frac{4}{c}
  \mathcal{F} \mathbf{H}] \,,
\end{equation}
%Eq.10
\begin{equation}
\nabla \times \mathbf{K}(\mathbf{x})=\frac{\zeta}{2\pi \varepsilon_0}  \nabla_{\mathbf{x}}\times \int
\frac{d^3y}{\left|\mathbf{x} - \mathbf{y} \right|}\nabla_{\mathbf{y}}\times [ 4 \mathcal{F} \mathbf{D} + \frac{7}{c}
 \mathcal{G} \mathbf{H}] \,. 
\end{equation}
To isolate the induced (non-linear) effects we define new electric and magnetic fields  with respect to the (linear) Maxwell theory as
%Eq.11
\begin{equation}
\mathcal{E}(\mathbf{x}) = \mathbf{E}(\mathbf{x}) - \frac{1}{\varepsilon_0} \mathbf{D}_M(\mathbf{x})\, ,
\end{equation}
%Eq.12
\begin{equation}
\mathcal{B}(\mathbf{x}) = \mathbf{B}(\mathbf{x}) - \mu_0 \mathbf{H}_M(\mathbf{x})\, ,
\end{equation}
with $\nabla \times \mathcal{E}(\mathbf{x}) = 0$, and $\nabla \cdot \mathcal{B}(\mathbf{x})= 0$. Using in Eqs.(11)-(12) the expressions for $\mathbf{E(\mathbf{x})}$ and $\mathbf{B(\mathbf{x})}$ that result from inverting Eqs.(5)-(6), together with Eqs.(7)-(10), it follows  to leading order in $\zeta$
%Eq.13
\begin{equation}
\mathcal{E}(\mathbf{x}) = \frac{\zeta}{2\pi \varepsilon_0^2}  \nabla_{\mathbf{x}} \int
    \frac{d^3y}{\left|\mathbf{x} - \mathbf{y} \right|}
    \nabla_{\mathbf{y}} \cdot \left[ 4 \mathcal{F}_M \mathbf{D}_M  +
    \frac{7}{c} \mathcal{G}_M  \mathbf{H}_M \right] \,,
\end{equation}
%Eq.14
\begin{eqnarray}
\mathcal{B}(\mathbf{x}) &=& \frac{\zeta}{2\pi \,\varepsilon_0^2\, c^2}\,  \nabla_{\mathbf{x}}\times  \int
    \frac{d^3y}{\left|\mathbf{x} - \mathbf{y} \right|}\,
    \nabla_{\mathbf{y}} \times \left[- 4 \mathcal{F}_M \mathbf{H}_M \right. \nonumber\\ [.3cm]
     &+& \left.
    7\,c\, \mathcal{G}_M  \mathbf{D}_M \right] \,.
\end{eqnarray}
The two equations above provide general analytical expressions to compute the induced fields in terms of the  Maxwell fields. The latter can be obtained as usual, once the external classical charge and/or current distributions are specified.\\

We now consider as a source an electrically neutral particle of radius $\underline{a}$, possessing a  magnetic dipole moment, in the presence of an external, quasistatic,  constant electric field $\mathbf{E}_0$. The details of how  this magnetic dipole moment is produced will play no role in the sequel. In fact, the final result for the induced electric dipole moment, generated by non-linearity in the presence of  $\mathbf{E}_0$, depends only on $\underline{a}$ and $\mathbf{m}$ (and obviously on $\mathbf{E}_0$). Hence, for simplicity, we consider
a  current density uniformly distributed on the surface of a sphere of radius $\underline{a}$
%Eq.15
\begin{equation}
    \mathbf{j}= \frac{3\, {|\mathbf{m}|}}{4 \pi a^3} \, \delta(r-a)\, \mathbf{\hat{e}_{\phi}}\,.
\end{equation}
This current distribution gives rise to a magnetic dipole-type field 
%Eq.16
\begin{equation}
    \mathbf{B}_d=\frac{\mu_0}{4 \pi}\left\{
      \frac{\left[
    3\left(\mathbf{{m}}\cdot \hat{\mathbf{e}}_r \right) \hat{\mathbf{e}}_r -
    {\mathbf{m}}
    \right]}{r^3} \,\Theta(r-a) + \frac{2 \,{\mathbf{m}}}{a^{3}} \,\Theta(a-r)
    \right\}\;,
\end{equation}
where $\mathbf{{m}}$ is identified with the magnetic dipole moment of the source.
A similar result is obtained for a uniformly magnetized sphere of the same radius $\underline{a}$. Since the central expressions, Eqs.(13)-(14), were derived assuming $E \equiv c B < E_c$, the following constraint follows
%Eq.17
\begin{equation}
\frac{|\mathbf{m}|}{r_c^3}< \frac{2 \pi {m_e}^2c^2}{\hbar\,e\,\mu_0}\,,
\end{equation}
where $r_c$ stands for a critical (minimum) radial distance from the source, e.g. if $|\mathbf{m}| \simeq 10^{-26} \mbox{A} \,\mbox{m}^2$, then it follows that $r_c > 10\, \mbox{fm}$.
We shall return to this bound later on in the conclusions. 
With $\mathbf{D}_M= \epsilon_0 \, \mathbf{E}_0\,,$ and 
$\mathbf{H}_M=\frac{1}{\mu_0}\, \mathbf{B}_d\,,$ the densities $\mathcal{F}_M$ and $\mathcal{G}_M$ become
%Eq.18
\begin{eqnarray}
    \mathcal{F}_M &=&\frac{1}{2}\, {\mathbf{E}_0}^2 -
    \frac{1}{2} \left( \frac{c\, \mu_0}{4 \pi} \right)^2 \left\{
    \frac{\left[ 3 \left( \mathbf{m} \cdot \hat{\mathbf{e}}_r \right)^2 
     +  \mathbf{m}^2 \right]}{r^6}
    \,\Theta(r-a) \right. \nonumber\\ [.3cm]
    &+& \left. \frac{4\, \mathbf{m}^2}{a^6}\,\Theta(a-r)\right\}\,,
\end{eqnarray}
%Eq.19
\begin{eqnarray}
    \mathcal{G}_M &=&\frac{c\, \mu_0}{4 \pi} \left\{\frac{
    \left[ 3 \left( \mathbf{m} \cdot \hat{\mathbf{e}}_r \right)
    \left( \mathbf{E}_0 \cdot \hat{\mathbf{e}}_r \right) -
    \left( \mathbf{m} \cdot {\mathbf{E}_0} \right) \right]}{r^3}
    \,\Theta(r-a) \right. \nonumber\\ [.3cm]
    &+& \left.  \frac{2\left( \mathbf{m} \cdot {\mathbf{E}_0} \right)}{a^3} \,\Theta(a-r) \right\}\,.
\end{eqnarray}

Choosing the magnetic dipole moment along the z-axis,  $\mathbf{m} = |\mathbf{m}|\hat{\mathbf{e}}_z$,
and the external electric field $\mathbf{E}_0$ in the x-z-plane, forming an angle $\psi$ with the z-axis,
%Eq.20
\begin{equation}\label{I12}
    \mathbf{E}_0= E_0 \left(\sin \psi \, \hat{\mathbf{e}}_x + \cos \psi\, \hat{\mathbf{e}}_z \right)\,.
\end{equation}
it follows that
%Eq.21
\begin{equation}
      \left( \mathbf{E}_0 \cdot \hat{\mathbf{e}}_r \right)= E_0\;(
    \cos\psi \, \cos\theta + \sin\psi \, \sin\theta\, \cos \phi)\,,
\end{equation}
where the angular function in parenthesis is cos $\gamma$, where $\gamma$ is the angle between the external electric field and the radial direction. Concentrating on the induced electric field, the integrand in Eq.(13) is given by
%Eq.22
\begin{eqnarray}
       &&   \nabla \cdot \left(4 \mathcal{F}_M \mathbf{D}_M
    + \frac{7}{c} \mathcal{G}_M \mathbf{H}_M \right)  =
    \frac{\mu_0 |\mathbf{m}|^2 |\mathbf{E_0}|}{(4 \pi)^2} \nonumber\\ [.3cm]
    &\times& \left\{\frac{6}{a^6}
    \left[ \cos\gamma \left(1+6 (\cos\theta)^2 \right)
    -7 \cos\psi \cos\theta
    \right] \delta(r-a)\right. \nonumber\\ [.3cm]
 &-&              \left.    \frac{9}{r^7} \left[\cos\gamma
     \left( 1+ 4 \left( \cos\theta \right)^2 \right)
    -\cos\psi \cos\theta \right] \right. \nonumber\\ [.3cm]
   &\times& \left. \Theta(r-a)\frac{}{} \right\}\,.
\end{eqnarray}
Substituing this expression in Eq.(13), expanding the inverse distance in terms of spherical harmonics,  using their orthogonality properties, and performing the three-dimensional integration, the final result for the induced electric field is
%Eq.23
\begin{equation}
    \begin{array}{c} \displaystyle
      \mathcal{E}(\mathbf{x})=-\frac{\zeta \mu_0 |\mathbf{m}|^2 |\mathbf{E_0}|}{8 \pi^2{\epsilon_0}^2} \, \nabla_x\ 
    \\ \\ \displaystyle
    \left\{ \frac{1}{ |\mathbf{x}|^2 a^3}
    \left[\frac{24}{5}
   \sqrt{3\pi} \cos\psi
   Y_{1\,0}(\theta ,\phi )+
    \right.\right.
    \\ \\ \displaystyle
    \left.\left.
    +\frac{13}{5} \sqrt{\frac{2\pi}{3}} \sin \psi  \left\{Y_{1\,1}(\theta ,\phi
   )-Y_{1\,-1}(\theta ,\phi )\right\}
    \right]+\right.
    \\  \\ \displaystyle
    \left.
    +\frac{1}{|\mathbf{x}|^5}\left[\cos \psi  \left(-\frac{4}{5} \sqrt{3 \pi }
   Y_{1\,0}(\theta ,\phi )-\frac{18}{5} \sqrt{\frac{\pi
   }{7}} Y_{3\,0}(\theta ,\phi )\right)+
    \right.\right.
    \\  \\  \displaystyle
    \left.\left.
    +\sin \psi \left(\frac{3}{5} \sqrt{\frac{3 \pi }{2}}
   \left\{Y_{1\,1}(\theta ,\phi )-Y_{1\,-1}(\theta ,\phi )\right\}+
    \right.\right.\right.
    \\ \\ \displaystyle
    \left.\left.\left.
    +\frac{6}{5} \sqrt{\frac{3
   \pi }{7}} \left\{Y_{3\,1}(\theta ,\phi )- Y_{3\,-1}(\theta ,\phi )\right\}\right)
    \right]
    \right\}\,,
    \end{array}
\end{equation}
where $|\mathbf{x}| = r$.
The first term in brackets above, with the $1/|\mathbf{x}|^2$ behaviour, corresponds to the induced electric dipole moment. For the case in which the external electric field $\mathbf{E_0}$ is parallel to the magnetic dipole moment along the z-axis, i.e. for $\psi=0$, the induced electric dipole field is given by
%Eq.24
\begin{equation}
      \mathcal{E}(\mathbf{x})
      = - \nabla_x
    \left\{ \frac{1}{4\pi \epsilon_0}\left(
    \frac{18 \zeta \mu_0 |\mathbf{m}|^2 |\mathbf{E_0}|}{5\pi a^3 \epsilon_0}\right) \frac{\hat{\mathbf{e}}_z \cdot\hat{\mathbf{e}}_r}{|\mathbf{x}|^2}
    \right\}\,.
\end{equation}
The induced electric dipole moment, located at the origin, is then given by
%Eq.25
\begin{equation}
    \mathbf{p}_{IND} = \left( \frac{18 \zeta \mu_0 |\mathbf{m}|^2 |\mathbf{E_0}|}{5\pi a^3 \epsilon_0}\right) \hat{\mathbf{e}}_z
\end{equation}
When $\mathbf{m}$ and $\mathbf{E}_0$ are not parallel, the term proportional to $\sin\psi$ in Eq.(23) leads to an explicit dependence on the  azimuthal angle $\phi$, viz.
%Eq.26
\begin{equation}
      \mathcal{E}(\mathbf{x})= - \nabla_x
        \left\{ \frac{1}{4\pi \epsilon_0} \,
    \frac{\mathbf{p}(\psi)|_{IND} \cdot  \hat{\mathbf{e}_r}}{|\mathbf{x}|^2}
    \right\}\,,
\end{equation}
where the induced electric dipole moment is now
%Eq.27
\begin{equation}
    \mathbf{p}(\psi)|_{IND}= \left( \frac{ \zeta \mu_0 |\mathbf{m}|^2 |\mathbf{E}_0|}{10\pi \epsilon_0 a^3}\right) \left[ 36 \frac{\mathbf{E}_0}{|\mathbf{E}_0|}  - 49
     \left( \frac{\mathbf{E}_0}{|\mathbf{E}_0|} \cdot \hat{\mathbf{e}}_x \right)  \hat{\mathbf{e}}_x \right]\,,
\end{equation}
which reduces to Eq.(25) when $\psi=0$. This induced electric field is of the electric dipole-type in its radial $1/|\mathbf{x}|^3$ dependence, but it has a manifestly peculiar angular dependence. For instance, along the z-axis, and unlike a standard electric dipole field, it has a non-zero component along $\mathbf{e}_\theta$ that depends on the azimuthal angle $\phi$. It also has a non-zero component along the direction of $\mathbf{e}_\phi$, as may be appreciated by writing the induced electric field  in spherical coordinates, i.e.
%Eq.28
\begin{eqnarray}
&&\mathcal{E}(\mathbf{x})= \;
   \frac{\zeta  \mu _0 |\mathbf{m}|^2 |\mathbf{E}_0| }{40 \pi^2 \epsilon _0^2 a^3 \mathbf{|x|}^3}
    \left\{\frac{}{}
    2\,  \left[\frac{}{}36
   \cos \theta \cos \psi  \right. \right.  \nonumber\\ [.3cm]
   &-& \left. \left. 13 \sin \theta \cos
   \phi  \sin \psi \frac{}{}\right]\,{\mathbf{\hat{e}}_r}+ 
  \left[\frac{}{}13 \cos
   \theta \cos \phi  \sin \psi \right. \right.  \nonumber\\ [.3cm]
    &+& \left. \left. 36 \sin \theta
   \cos \psi \frac{}{}\right] {\mathbf{\hat{e}}_\theta} -13\,  \sin \phi \sin
   \psi \;{\mathbf{\hat{e}}_\phi}\frac{}{}
    \right\}
\end{eqnarray}
This feature should help to differentiate this induced effect from  other predictions of an intrinsic electric dipole moment of the neutron in extensions of the Standard Model \cite{EDM_ST}. To arrive at an estimate of the size of the induced electric dipole moment of the neutron, we consider first the case $\psi = 0$, i.e. the external electric field $\mathbf{E}_0$ parallel to the z-axis, so that the induced electric dipole field has the usual angular dependence. Expressing $|\mathbf{E}_0|$ in V/m, and the radius $\underline{a}$ in fermi, Eq.(25) gives
%Eq.29
\begin{equation} 
 |\mathbf{p}|_{IND} \simeq 10^{-33}\;|e|\, \mbox{cm}  \left[\frac{|\mathbf{E}_0|(\mbox{V/m})}{a(\mbox{fm})^3}\right]\;.
\end{equation}
A reasonable estimate of the neutron radius can be obtained from its mean squared radius \cite{PDG} in which
case $a \simeq 0.35 \,\mbox{fm}$ . Current experimental sensitivity \cite{BAKER} is at the level of $10^{-26} |e| \mbox{cm}$, which could improve by two orders of magnitude in the SNS experiment at Los Alamos \cite{LA}, in the CryoEDM experiment at ILL  \cite{Cryo}, or in the nEDM experiment at PSI \cite{nEDM}. However, current experimental setups involve an external magnetic field, around which the neutron spin precesses, and an external electric field to probe the electric dipole moment. This external electric field, though, is at present well below the one  required to bring $|\mathbf{p}|_{IND}$ to an observable level. In fact, one would need fields as high as $|\mathbf{E}_0| \simeq 10^{7} - 10^{8}(\mbox{V/m})$. Such intense fields are indeed present in some crystals at the level of $|\mathbf{E}_0| \simeq 10^{10} \mbox{V/m}$, but would require a completely different setup \cite{XTAL}. In this case, with $a \simeq 1 \mbox{fm}$, the order of magnitude of the induced electric dipole moment would be $|\mathbf{p}|_{IND} \simeq 10^{-23}\, |e| \mbox{cm}$. While this range is within current sensitivity, it is crucial to change the experimental setup so that the external electric field is as high as indicated above.
Turning to the more general case $\psi \neq 0$, the order of magnitude of $\mathbf{p}|_{IND}$ does not change much, but the peculiar angular dependence of the electric field offers some hope of improving the chance of observability. To aid in the design of the experiment we 
point out that, from Eq.(27), the induced electric dipole moment has a component along the direction of $\mathbf{E}_0$, and a component along the x-axis. Hence it will interact with $\mathbf{E}_0$ and change the Larmor frequency of the neutron magnetic moment around the external magnetic field.\\
In connection with the bound, Eq.(17), which follows from the weak field approximation, it would be  satisfied quite easily for the neutron if the experiment is performed using a crystal, since in this case $r_c >> 10 fm$. 
We mention in closing that we have computed the induced magnetic field from Eq.(14). The leading term in this field is of the magnetic dipole type, and thus it is a correction to the field produced by the source (neutron). However, its magnitude is many orders of magnitude smaller than the latter, so that it can safely be ignored.\\
The authors wish to thank Harald Griesshammer, Ulli K\"{o}ster, and Oliver Zimmer for valuable discussions. This work has been supported in part by NRF (South Africa), FONDECYT 1060653, 1095217 and 7080120 (Chile), Centro de Estudios Subatomicos (Chile), CONICET PIP 01787, and UNLP Proy. 11/X492 (Argentina).

%***********************************************************************************

\end{document}